\documentclass[lettersize,journal]{IEEEtran}

\usepackage{amsmath,amsfonts}
\usepackage{algorithmic}
\usepackage{algorithm}
\usepackage{array}
\usepackage{textcomp}
\usepackage{stfloats}
\usepackage{url}
\usepackage{verbatim}
\usepackage{graphicx}
\usepackage{gensymb}
\usepackage{cite}
\usepackage{xcolor}
\usepackage{booktabs} 
\usepackage{makecell}
\usepackage{subcaption}
\usepackage{graphicx}
\usepackage{multirow}
\usepackage{float} % 如果需要使用 [H] 固定位置
\usepackage[colorlinks=true, linkcolor=blue, citecolor=blue, urlcolor=blue]{hyperref}
\begin{document}

\title{A Multi-Scale ResNet-augmented Fourier Neural Operator Framework for High-Frequency Sequence-to-Sequence Prediction of Magnetic Hysteresis}

\author{
    Ziqing~Guo, 
    Xiaobing~Shen, 
    and~Ruth~V.~Sabariego,~\IEEEmembership{Member,~IEEE}%
\thanks{Manuscript received Month Day, 2026; revised Month Day, 202X. (Corresponding author: Ziqing Guo.)}%
\thanks{Ziqing Guo and Ruth V. Sabariego are with the Department of Electrical Engineering (ESAT), KU Leuven, 3001, Leuven, Belgium, and also with EnergyVille, 3600, Genk, Belgium (e-mail: ziqing.guo@kuleuven.be; ruth.sabariego@kuleuven.be).}%
\thanks{Xiaobing Shen is with the Department of Electrical Engineering, Xiamen University of Technology, 361024, Xiamen, China (e-mail: xbing.shen@xmut.edu.cn).}%
}
% The paper headers
\markboth{Journal of \LaTeX\ Class Files,~Vol.~14, No.~8, August~2021}%
{Shell \MakeLowercase{\textit{et al.}}: A Sample Article Using IEEEtran.cls for IEEE Journals}

% \IEEEpubid{~\copyright~2021 IEEE}
% Remember, if you use this you must call \IEEEpubidadjcol in the second
% column for its text to clear the IEEEpubid mark.

\maketitle

\begin{abstract}
Accurate modeling of magnetic hysteresis is essential for high-fidelity power electronics device simulations. The transient hysteresis phenomena such as the ringing effect and the minor loops are the bottleneck for the accurate hysteresis modeling and the core losses estimation. To capture the hysteresis loops with both the macro structure and the micro transient details, in this paper, we propose the multi-scale ResNet augmented Fourier Neural Operator (Res-FNO).
The framework employs a hybrid input structure that combines sequential time-series data with scalar material labels through specialized feature engineering. Specifically, the time derivative of magnetic flux density ($\frac{dB}{dt}$) is incorporated as a critical physical feature to enhance the model sensitivity to high-frequency oscillations and minor loop triggers. The proposed architecture synergizes global spectral modeling with localized refinement by integrating a multi-scale ResNet path in parallel with the FNO blocks. This design allows the global operator path to capture the underlying physical evolution while the local refinement path, compensates for spectral bias and reconstructs fine-grained temporal details. Extensive experimental validation across diverse magnetic materials from 79 to Material 3C90 demonstrates the strong generalization capability of the proposed Res-FNO, proving its robust ability to model complex ringing effects and minor loops in realistic power electronic applications.
\end{abstract}


\section{Introduction}
\label{sec:intro}
\IEEEPARstart{M}{agnetic} components, such as inductors and transformers, are indispensable in nearly all power electronic system, yet they often dominate volume, weight, and losses, limiting overall system performance and efficiency \cite{sullivan2016small, how_magnet_2023, serrano_why_magnet, dang_multi_stage}. Accurate modeling of the magnetic hysteresis loops of power magnetic materials is therefore essential for reliable design, optimization and simulation of high-efficiency, high-power-density converters \cite{muehlethaler2011loss}. However, the underlying physics is extremely complex: the material response depends nonlinearly in coupled manner on a multitude of factors, including excitation frequency, peak flux density, DC bias, temperature, core geometry, transient history and waveform shapes such as sinusoidal, trapezoidal and pulse width modulation (PWM) ect. \cite{kacki2019study}. Traditional empirical models such as the Steinmetz equation (SE) \cite{steinmetz1984law} and its generalizations, including the improved generalized Steinmetz equation (iGSE) \cite{venkatachalam2002accurate} and the improved-improved generalized Steinmetz equation (i$^2$GSE) \cite{muhlethaler2011improved}, are limited to narrow operating regimes and cannot simultaneous capture all these intertwined effects, while physics-based models, such as Jiles-Atherton (J-A) model \cite{jiles1986theory}, Preisach model \cite{classical_Preisach_model} and the energy-based model \cite{energy_based_model} are computationally expensive or inaccurate under realistic, nonsinusoidal, high-frequency excitations typical of modern power electronics.\\

With recent advances in computational intelligence, data-driven models have demonstrated superior flexibility and accuracy \cite{li2022framework} in the magnetic characteristics modeling. Research in this field primarily branches into two categories. The first is scalar-to-scalar modeling, which extracts excitation features such as peak values, frequency, DC bias, waveform hot vector, etc. and environmental factors like temperature to directly predict core loss \cite{XB_iron_loss, Magnetic_core_loss_NN}. While effective for specific cases, these ``black-box'' models cannot reconstruct the underlying hysteretic trajectories, which limits their predictive accuracy and physical consistency \cite{Hardcore}. To bridge this gap, sequence-to-sequence (seq-to-seq) models \cite{sutskever2014sequence, serrano2022compel} have been developed to map flux density ($B$) waveforms directly to magnetic field strength ($H$) waveforms. Hybrid Preisach-recurrent networks and standalone deep neural networks (NNs) model arbitrary hysteresis processes with high fidelity \cite{preisach_recurrent_NN, quondam2022neural} and physics-aware recurrent neural networks incorporate history dependence and generalize to first-order reversal curves and minor loops \cite{chandra2025generalizable}. Neural operators, which include DeepONet, Fourier neural operator (FNO), have been applied to learn operator mappings between magnetic fields, enabling rate-independent prediction of novel hysteresis curves not seen in training \cite{dynamic_hysteresis_zguo}; and rate-independent FNO variants have shown strong extrapolation capability for complex minor-loop trajectories \cite{chandra_rate_independent}.\\ 

Recently, Significant progress has been made in high-frequency seq-to-seq $B-H$ prediction, largely enabled by the open-source MagNet database, which provides more than $500000$ experimentally measured $B-H$ loops across a wide range of frequency, flux densities, waveforms, DC biases, and temperatures for multiple materials \cite{Magnet_challenge}. Li et al. introduced the MagNet framework and demonstrated seq-to-seq LSTM encoder–decoder architectures for full $B–H$ loop prediction, mapping $B(t)$ to $H(t)$, showing that NNs can serve as ``active datasheets" with superior accuracy and generality compared with analytical models \cite{magnet_AI_haoran_li}. Serrano et al. further extended this data-driven paradigm by developing advanced NN models that can capture the complex nonlinearities of magnetic materials across diverse operating conditions, effectively bridging the gap between massive experimental data and practical power electronics design \cite{serrano_why_magnet}. Subsequent works have extended these ideas: \cite{transformer_encoder_decoder} proposed a Transformer-based encoder-projector-decoder architecture that leverages attention mechanisms \cite{vaswani2017attention} to capture temporal dependencies and scalar operating conditions. Similarly, the top-performing models in the MagNet Challenge 2023, particularly the HARDCORE framework \cite{Hardcore}, established comprehensive benchmarks incorporating extensive data preprocessing and convulutional neural network (CNN)-based architectures to evaluate performance across diverse magnetic materials. Despite their success, these models typically rely on substantial training datasets to ensure high accuracy, and their performance tends to degrade in data-constrained scenarios. Furthermore, since these approaches primarily focus on core loss estimation, the predicted hysteresis loops often lack sufficient precision and fail to accurately capture transient phenomena. \cite{Huang2025TPEL} introduced a hysteresis model based on hysteresis separation theory, which reduces the training data requirement, yet still necessitates a considerable amount of samples. \cite{dynamic_hysteresis_zguo} investigated various neural operators for hysteresis modeling and identified the Fourier Neural Operator (FNO) as exhibiting strong generalization capability for dynamic hysteresis, which can achieve great accuracy with a small amount of training data. However, their study was limited to sinusoidal excitations with frequencies below 1000 Hz, leaving the model's performance under complex non-sinusoidal waveforms and higher frequency regimes unexplored.

To overcome these limitations, a multi-scale ResNet-augmented Fourier Neural Operator (Res-FNO) specifically designed to mitigate spectral bias in high-frequency sequence-to-sequence B-H hysteresis modeling is proposed. The core innovation lies in the synergistic integration of a Fourier neural operator \cite{li2023fourier}, which efficiently captures global frequency-domain interactions across the entire B–H trajectory, with multi-scale ResNet \cite{resnet2016}branches that explicitly emphasize high-frequency residuals and local sharp transitions. The proposed architecture overcomes the low-frequency bias of vanilla neural operators, long-short term memory, Transformer models, and standard recurrent networks, achieving superior accuracy in predicting full $B–H$ sequences under wide range of excitations and arbitrary waveforms. This framework is highly generalizable across materials and operating conditions while maintaining a compact parameter count, making it practical for real-time circuit simulation and hardware-in-the-loop applications.
The main contributions of this article are as follows:
\begin{itemize}
    \item To improve sensitivity to transient phenomena such as ringing and minor loops, a hybrid input strategy with the multi-input processing is introduced, incorporating the time derivative $dB/dt$ and scalar operating conditions. \\
 
    \item A multi-scale Res-FNO framework is proposed for high-frequency sequence-to-sequence magnetic hysteresis modeling with the transient details. \\
   
    \item Results demonstrate strong generalization capability across diverse magnetic materials from the MagNet Challenges. Particularly on transient and complex minor-loop trajectories, the proposed model achieves superior accuracy in both global loop shape and local transient details compared to baseline FNO models.\\    
\end{itemize}

The remainder of this paper is structured as follows. Section \ref{sec: problem statement} describes the datasets and materials used in this study. Section \ref{sec: model description} presents the proposed multi-scale Res-FNO architecture and its data feature engineering. Section \ref{sec: results} details the experimental validation, ablation studies, and generalization results across diverse materials and minor-loop scenarios. The conclusion is summarized in Section \ref{sec:conclusion}.

\section{Problem formulation and data representation}
\label{sec: problem statement}
While existing literature primarily focuses on the direct approximation of scalar core loss density,
This work purses to modeling the dynamic hysteresis operator $\mathcal{P}: B(t) \mapsto H(t)$ by NNs. By formulating the problem as a seq-to-seq mapping, the dynamic hysteresis trajectory can be reconstructed, and then the core loss density can subsequently derived via the periodic integration of the predicted $B-H$ loop. This trajectory-based approach offers two significant advantages. First, it enables the high-fidelity capture of intricate hysteresis characteristics, such as frequency-dependent widening and nonlinear saturation effects. Second, beyond its utility as a high-precision loss post-processing tool, the proposed model can be integrated into the Finite element method framework as a differential surrogate hysteresis model \cite{Hardcore}.\\
The experimental data utilized in this study are primarily sourced from the MagNet Challenge 2023. The dataset was systematically structured into two tiers with increasing complexity to facilitate a robust assessment of model performance under progressively challenging extrapolation scenarios. The first tier encompasses ten material types characterized by relatively stable magnetic behaviors and balanced data distributions. In contrast, the second tier, comprising materials 3C92, T37, 3C95, 79, and ML95S, presents significant modeling hurdles, including concurrent high-frequency, high-peak excitations, sparse training samples, and missing data under specific operational conditions in training data. In addition, these selected materials span a diverse range of physical characteristics: 
\begin{itemize}
    \item The power ferrites, including 3C92, 3C95 and ML95S, which are engineered for minimized core loss in high frequency power converters.\\
    \item High permeability material, represented by T37, primarily utilized in EMI filters for noise suppression. \\
    \item High Nickel alloys, represented by 79, which exhibits distinct magnetic properties and energy dissipation mechanisms compared to standard ferrites.\\
\end{itemize}
A wide range of operation points were measured for each material, covering sinusoidal, trapezoidal and square waveforms, as well as ringing effect due to a high switching speed of the used semiconductors. The frequency varies from 50 to 800 kHz, the temperature from 25 $\celsius$ to 90 $\celsius$. An in-depth description of the dataset, the occurring waveforms, the data acquisition process, the lab setup, and the data quality control can be found in \cite{Hardcore, Magnet_challenge}.\\ 
While models proposed for the MagNet Challenge 2023 have demonstrated competitive performance on the first tier, they often struggle with the intricate nonlinearities and rigorous extrapolation requirements of the second tier. Consequently, this work focuses on the second-tier materials to test the accuracy and the generalizability of the proposed architecture. Furthermore, to specifically evaluate the capability of proposed model in capturing the minor loop behavior, we incorporate additional data for material 3C90 sourced from the MagNet Challenge 2. This inclusion provides a diverse set of complex minor loop trajectories, further ensuring the generalizability of the model across both global evolution and local hysteretic details.

\begin{figure}
    \centering
    \includegraphics[width=1.0\linewidth]{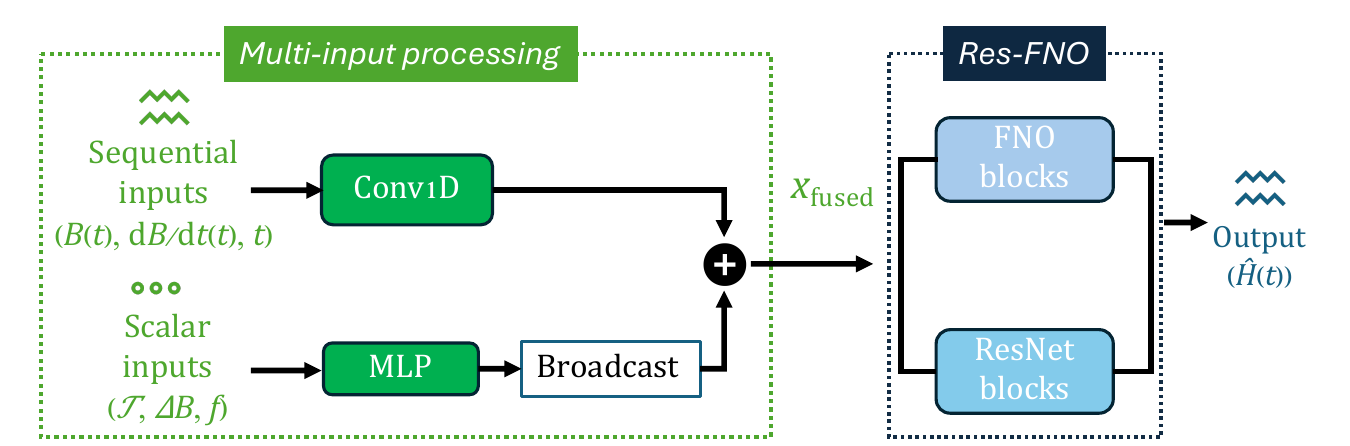}
    \caption{Structure of multi-scale Res-FNO: (Left) Multi-input Processing to fuse the scalar and sequential inputs; (Right) Res-FNO with parallel FNO blocks and ResNet blocks.}
    \label{fig:res_fno_structure}
\end{figure}

\begin{figure}
    \centering
    \includegraphics[width=0.7\linewidth]{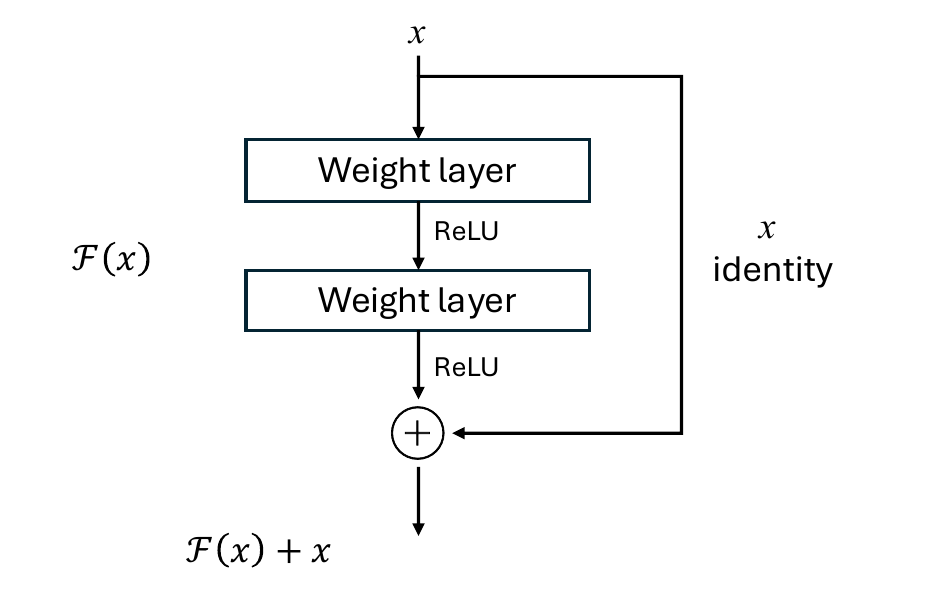}
    \caption{Residual learning: a building block \cite{resnet2016}}
    \label{fig:resnet_block}
\end{figure}

\section{Model description}
\label{sec: model description}
In this paper, the hybrid multi-scale ResNet augmented Fourier Neural Operator (Res-FNO) is proposed for high frequency magnetic hysteresis modeling. As shown in Fig. \ref{fig:res_fno_structure}, there are two parts in this model, the multi-input processing stage for fusing the multiple inputs, and the Res-FNO backbone with the parallel FNO and ResNet blocks.

\subsection{Multi-input processing}
\label{sec: multi input processing}
To effectively capture the disparate features of temporal sequences and scalar features, a dual stream processing architecture is employed as shown in Fig. \ref{fig:res_fno_structure}. For the sequential data stream, including magnetic flux density $B(t)$, its temporal derivative $dB/dt$, and the time vector $t$, are fed into a one-dimensional convolutional layer (Conv1D). This Conv1D serves as a local feature extractor, identifying transient patterns and gradient transitions within the input sequence while mapping the raw data into a latent feature space. Simultaneously, the scalar parameter stream, consisting of environmental and operational constraints such as temperature ($\mathcal{T}$), frequency ($f$) and the peak-to-peak magnetic flux density ($\Delta B$), is processed through a dedicated Multi-layer perception (MLP). The use of $\Delta B$ instead of a simple peak value ensures a more precise representation of the magnetic flux excursion, especially in cases of asymmetric excitation where the flux is not centered around zero. This MLP encodes the global physical conditions into a high-dimensional feature vector that matches the embedding dimension of the sequential stream. To unify these multi-model features, after the MLP, the encoded scalar feature vector is spatially broadcast across the entire temporal dimension of the sequential stream, ensuring that global physical constraints are consistently represented at every time step. These two streams are then merged via an element-wise addition operation, resulting in a unified latent tensor that encapsulates both local dynamic evolution and global physical contexts. This fused representation serves as the comprehensive input for the subsequent Res-FNO backbone, ensuring that the model remains sensitive to external operating conditions during the spectral and temporal refinement stages.

\subsection{Res-FNO}
\label{sec:TFNO}
In this section, we detail the internal architecture of the proposed multi-scale Res-FNO, which serves as the central computational engine of the proposed framework.

\subsubsection{ResNet with the residual connection}
\label{sec:Resnet}
In the early evolution of deep convolutional neural networks (DCNNs), it was widely believed that increasing network depth would inherently enhance feature extraction capabilities. However, experimental evidence revealed that as the depth exceed a certain threshold, the accuracy of the training set would saturate and then degrade rapidly. It is crucial to distinguish this from vanishing gradients. In modern architectures, batch normalization \cite{batch_normalization} and proper weight initialization \cite{weight_initialization} have largely stabilized the gradient flow. Instead, the degradation suggests that approximating an identity mapping through multiple nonlinear layers is fundamentally difficult for numerical optimization. To address this challenge, He et.al \cite{resnet2016} introduced the residual learning framework. Rather than expecting a stack of layers to directly fit a desired underlying mapping $G(x)$, the framework explicitly lets these layers fit a residual mapping defined as $F(x) := H(x) - x$. Consequently, the original mapping is recast into:
\begin{equation}
    H(x) = F(x) + x.
\end{equation}

This concept is physically implemented through shortcut connections (also referred as skip connections). As illustrated in Fig. \ref{fig:resnet_block}, a typical residual building block consists of two primary paths:
\begin{itemize}
    \item Residual path: A series of weight layers, normalized layers, and activation functions (e.g. ReLU) that learn the incremental changes of the features. 
    \item A shortcut that bypasses the nonlinear transformations and propagates the input $x$ directly to the output.
\end{itemize}

The mathematical formulation of a basic block reads
\begin{equation}
    y = g(F(x, \{W_i\}) + x),
\end{equation}
where $y$ is the output vector and $g$ denotes the activation function. The integration of residual mechanisms provides several key benefits for optimizing deep models. Firstly, the shortcut connections create a ``highway'' for backpropagation, allowing the gradient of the loss function to bypass complex nonlinear layers and reach shallower layers with minimal attenuation. Secondly, research indicates that residual connections significantly reduce the complexity of the loss landscape, making the optimization process more stable and enabling faster convergence to a global optimum. Lastly, by preserving original information and learning only the refined increments, the network enhances its ability to reuse low level features throughout deeper structures, which is the main motivation we chose to use ResNet in our model.\\

\subsubsection{FNO block}
\label{sec:FNO block}
Neural operators were proposed to learn mappings between two infinite-dimensional function spaces based on a finite set of observed input-output pairs. Among various neural operator architectures, the Fourier Neural Operator (FNO) has demonstrated superior capacity and generalization performance in modeling hysteresis, as evidenced by the work of Guo et al. \cite{dynamic_hysteresis_zguo}. The FNO specifically parameterizes the integral kernel in Fourier space, allowing it to learn Fourier coefficients  directly from data \cite{li2020fourier}. In practice, FNO discretizes both the input $B(t)$ and output $H(t)$ on a uniform mesh. Each FNO block consists of two parallel paths as: A spectral branch and a skip connection as the illustrated ``FNO block" in Fig. \ref{fig:res_fno_block}. The spectral branch transforms the temporal (or spatial) input into the frequency domain via the Fast Fourier Transform (FFT), followed by a truncation of higher-order modes, a hyperparameter adjusted according to the complexity of the problems. The remaining frequency modes are multiplied by a learnable weight matrix. The processed information is then projected back to the original domain using the Inverse FFT (IFFT). Parallel to this, a skip connection is employed to preserve global information and mitigate the loss of critical features due to frequency truncation. The outputs from both paths are summed element-wise and passed through a nonlinear activation function to enhance the representational capacity of the model. For a more exhaustive mathematical derivation and parameter sensitivity analysis of the FNO layers, we refer the reader to \cite{dynamic_hysteresis_zguo}.

\subsubsection{FNO with Resnet}
\label{sec:FNO with resnet}
\begin{figure*}
    \centering
    \includegraphics[width=1.0\linewidth]{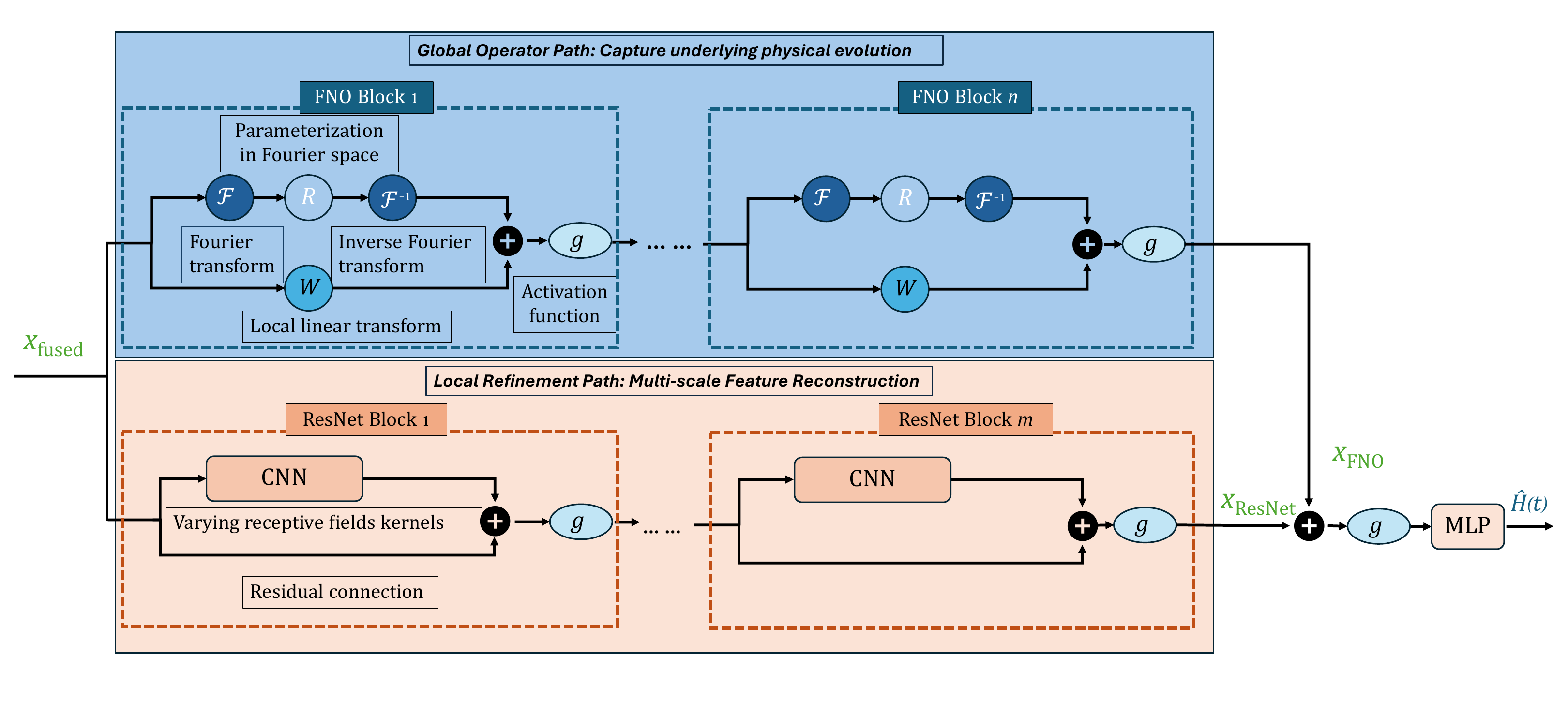}
    \caption{Structure of Res-FNO block: {Up} Global operator path with FNO blocks to capture underlying physical evolution; (Down) Local refinement path with ResNet blocks for multi-scale feature reconstruction.}
    \label{fig:res_fno_block}
\end{figure*}

While the FNO excels at learning global operators and capturing the underlying physical envelope in frequency space, it inherently suffers from spectral bias, where the network prioritizes the optimization of low frequency components \cite{multi_scale_fno}. In addition, due to the necessary truncation of Fourier modes, the spectral convolution acts as a low frequency filter, which inevitably smooths out sharp local transients. Critically, even if the number of retained modes is significantly increased, the FNO struggles to capture high frequency details, such as the ringing effect causing oscillations in hysteresis loops. In this paper, To synergize global spectral modeling with localized feature resolution, we proposed a multi-scale topology integrating FNO with the ResNet, as illustrated in Fig. \ref{fig:res_fno_block}. This design is motivated by the complementary strengths and inherent limitations of these two paradigms.\\
% \subsubsubsection{Global operator path: spectral evolution}
Firstly, the fused inputs $x_{fused}$ from the multi-input processing are fed into the global operator path, which consists of a sequence of $n$ FNO blocks. The structure of the FNO blocks are kept the same as in \cite{li2020fourier}, with the spectral processing path complemented by a local linear transform. By parameterizing the integral kernel in Fourier space, this path captures the long range physical evolution and the macro scale backbone of the hysteresis loop. Then, the special design of this model, the parallel local refinement path is introduced to compensate for the information loss incurred by special bias. This path comprises $m$ ResNet blocks specifically engineered to resolve localized nonlinearities. Each block utilizes CNN with varying receptive field kernels size ($k$). This multi-scale approach allows the model to patch the gaps left by mode truncation, effectively reconstructing high-frequency transients and fine-grained oscillations that the global operator might smooth out. 
The final output is synthesized through a multi-component additive fusion. The latent representation from the Global Operator Path ($x_{FNO}$) and the Local Refinement Path ($x_{ResNet}$) are combined as in Eq. \eqref{eq:output_res_fno}. Finally, the fused features are passed through a Multi-Layer Perceptron (MLP) to project the latent multi-scale representation back to the target physical space, yielding the predicted magnetic field strength $\hat{H}(t)$. 
\begin{equation}
\label{eq:output_res_fno}
\hat{H}(t) = \text{MLP} \left( g( \mathbf{x}_{FNO} + \mathbf{x}_{ResNet} ) \right)
\end{equation}
This integration ensures that the model maintains the global physical trend while preserving localized precision.

%%%%%%%%%%%%%%%%%%%%%%%%%%%%%%%%%%%%%%%%%%%%%%%%%%%
\subsection{Feature engineering}
\label{sec: feature engineering}
Hysteresis is inherently context-dependent, the same $B(t)$ input will yield different $H(t)$ responses depending on the operating frequency $f$ and temperature $\mathcal{T}$. So, besides the $H(t)$ sequence, the scalar features should also be included in the model. 
To address this, ``Feature channel engineering'' is employed. As shown in Fig.~\ref{fig:res_fno_structure} and in section \ref{sec: multi input processing}, the scalar features are mapped to the same dimension as the latent sequential features and integrated into  temporal signals. 
To mitigate the challenges of vanishing and exploding gradients in neural network, a multi-stage normalization pipeline is implemented:
\begin{itemize}
\item \textbf{Min-Max Scaling}: All input features, including the magnetic flux density sequence $B(t)$, the magnetic field intensity sequence $H(t)$, the time derivatives of $B$ ($\frac{dB}{dt}$) and the scalar operating conditions (frequency $f$, temperature $\mathcal{T}$ and the peak-to-peak value of $B$ ($\Delta B$))are normalized to the range $[-1, 1]$ using Min-Max scaling:
\begin{equation}
\label{eq:min_max}
x_{\text{norm}} =\frac{x - x_{\text{min}}}{x_{\text{max}} - x_{\text{min}}},
\end{equation}
where $x$ represents the original value, $x_{\text{min}}$ and $x_{\text{max}}$ are the minimum and maximum values of that feature across the training set, and $x_{\text{norm}}$ is the normalized value. This normalization ensures that the spectral convolutions in the FNO operate on a consistent numerical range and prevents features with larger magnitudes from dominating the learning process.

% \item \textbf{Logarithmic Compression}: Given the several orders of magnitude difference in loss density ($P_{loss}$), a logarithmic transform is applied to $P_{loss}$: $\tilde{P_{loss}} = \ln{P_{loss}}$. This is followed by Min-Max scaling to compress the heavy-tailed distribution of the excitation parameters.
% \item \textbf{Input Desensitization}: To prevent the model from overfitting to the grid resolution of the time-steps, we explicitly provide normalized time indices as an additional input channel, enhancing the model's ability to generalize to different sampling rates. 
% Specifically, in this work, the original waveforms consisting of 1024 time steps are downsampled to 205 points. This strategy significantly reduces the computational overhead while maintaining high prediction accuracy.
\item \textbf{Grid-Invariance and Resolution Robustness}: One of the core advantages of the FNO architecture is its theoretical resolution independence, which allows the model to learn operators between function spaces rather than discrete mappings. In this work, we leverage this property by employing a downsampling strategy on the temporal sequences to reduce computational overhead. As established in prior studies, the spectral-domain integral kernel in FNO provides sufficient resistance to variations in grid density. This ensures that the model can maintain high prediction fidelity and robust feature extraction even when operating at a coarser sampling rate than the original training data. Consequently, this allows for accelerated training on lower resolution data without sacrificing the accuracy of high resolution inference \cite{dynamic_hysteresis_zguo}. So, in our study, every groups of data are dowmsampled into less time points to speed up the training while keeping the accuracy. More details are provided in Section \ref{sec: results}.
\end{itemize}

\subsection{Loss function constraints}
\label{sec:Loss function}
Since the goal of the proposed multi-scale Res-FNO is for sequence to sequence model to predict the corresponding $H$, the loss function is defined as the point-by-point deviation between the predicted magnetic field $\hat{H}(t)$ and the experiment ground truth $H(t)$, defined as mean square error (MSE):
\begin{equation} 
\label{eq:loss_1}
\mathcal{L} = \frac{1}{M} \sum_{i=1}^{M} \left( \sum_{t=1}^{N} | H_{i}(t) - \hat{H}_{i}(t) |^2 \right), 
\end{equation}
where $M$ and $N$ denote the batch size and the number of time steps per period respectively. The MSE loss is chosen for its sensitivity to large derivations, which ensures the model effectively captures critical $H$ features such as peak values and rapid transitions.

\begin{figure*}
    \centering
    \includegraphics[width=0.9\linewidth]{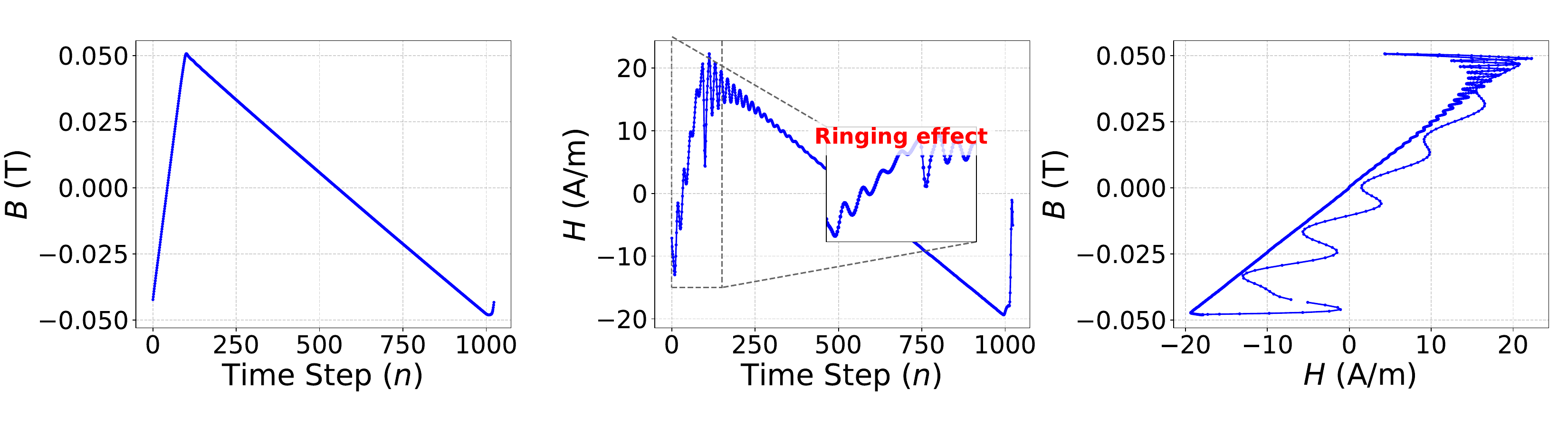}
    \caption{Typical magnetic characteristics under high-frequency excitation: (Left) Temporal waveform of the magnetic induction $B(t)$; (Middle) Corresponding magnetic field strength $H(t)$ exhibiting a pronounced ringing effect; (Right) The resulting $B$-$H$ hysteresis loops, illustrating the impact of temporal oscillations on the magnetic trajectory.}
    \label{fig:Ringing effect showing}
\end{figure*}

\begin{figure}
    \centering
    \includegraphics[width=1.0\linewidth]{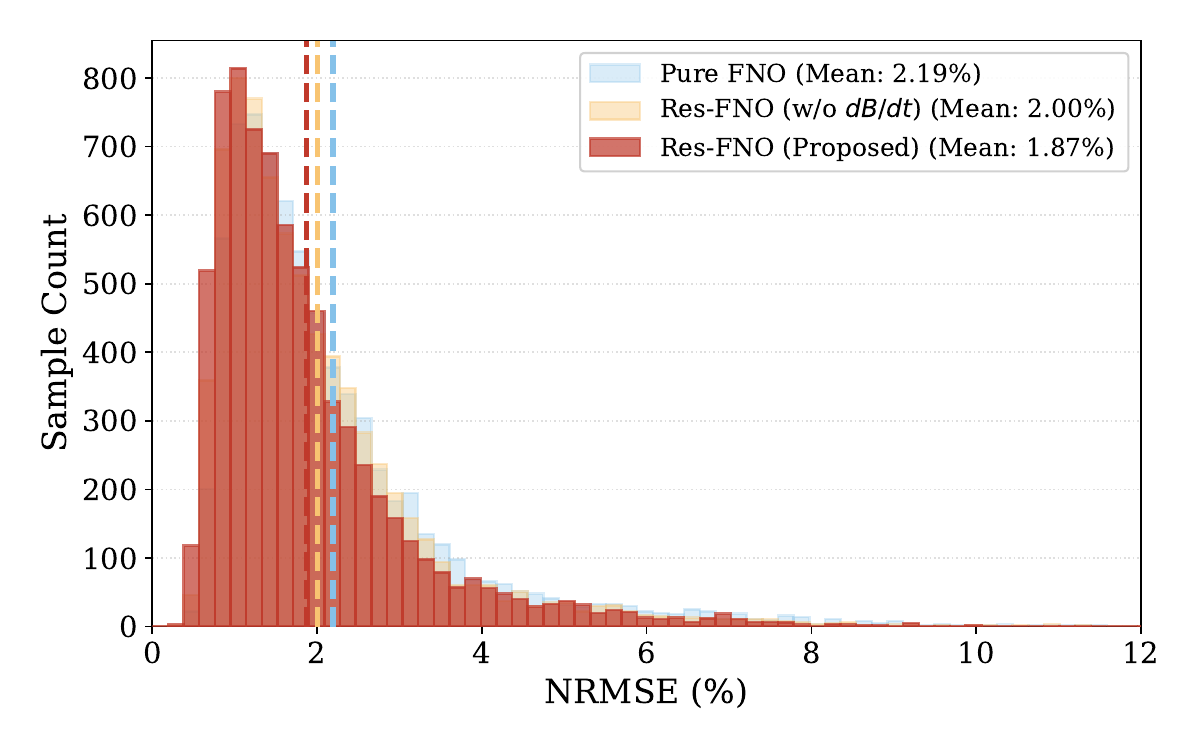}
    \caption{NRMSE distribution analysis for different model architectures.}
    \label{fig:NRMSE_comparison}
\end{figure}

\begin{figure}
    \centering
    \includegraphics[width=1.0\linewidth]{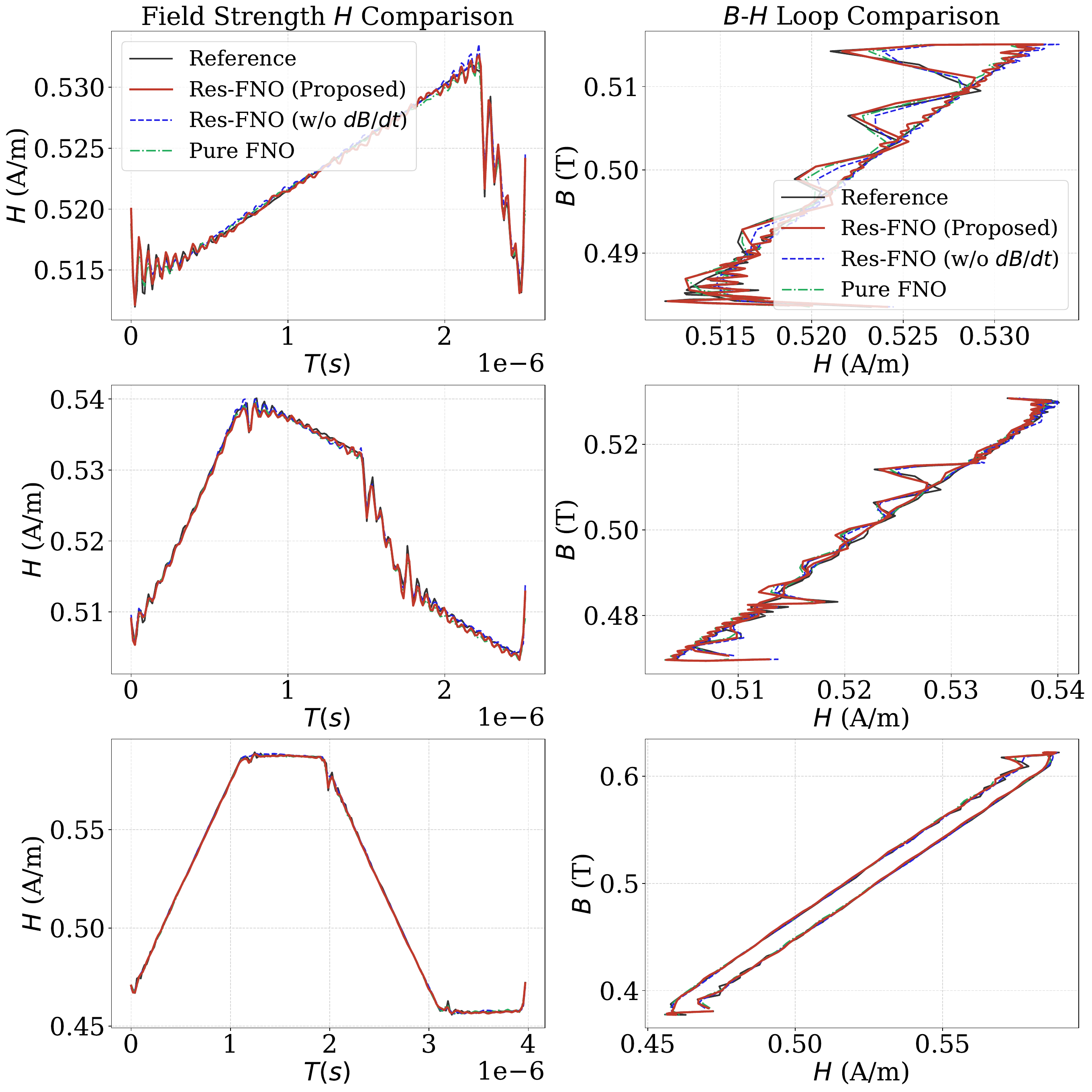}
    \caption{Comparison of predicted magnetic characteristics for using different model architectures: (Left) Temporal waveforms of the magnetic field strength $H(t)$; (Right) Resulting $B$-$H$ hysteresis loops.}
    \label{fig:Hysteresis_loops_M79}
\end{figure}
\section{Results}
\label{sec: results}
\subsection{Experimental configuration}
\label{sec:hardware}

\begin{figure}
    \centering
    \includegraphics[width=1.0\linewidth]{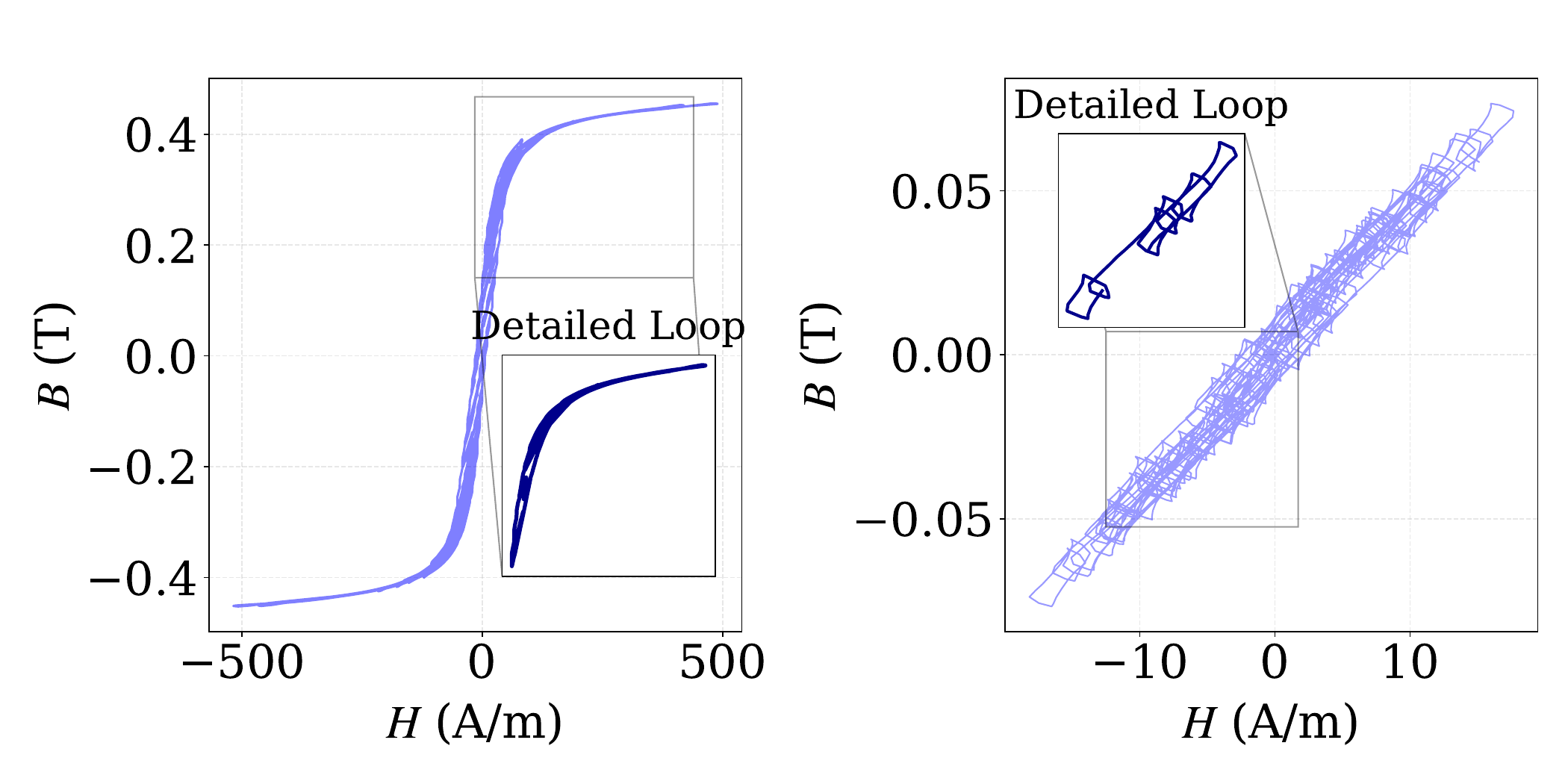}
    \caption{$B-H$ loops for 3C90 ferrite material under different excitation frequencies. (left:) under $f$ = 50 kHz, (right:) $f=800$ kHz}
    \label{fig:minor_loops_showing}
\end{figure}

The computational experiments were conducted on an Apple M1 processor using JAX and Equinox  framework. Unlike the traditional deep learning libraries, JAX utilizes accelerated linear algebra (XLA) to compile high-level Python and NumPy functions into highly optimized machine code, significantly enhancing execution speed on CPU architectures.\\
To evaluate the performance of model under extreme conditions, Material 79 from the second tier of the MagNet Challenge 2023 was selected as the primary benchmark for architectural evolution and hyperparameter tuning. This material is characterized by highly nonlinear dynamics and the smallest size of training data, leading to the highest predictive errors reported among various competitive models in the challenge \cite{Magnet_challenge}. In the original data provided by MagNet Challenge 2023, there are standardized train and test partitions specifically designed to evaluate the diverse predictive capabilities of neural networks. In our study, we strictly adhered to the official train and test datasets, but we further partitioned the train data provided into training and validation sets by ratio as $9:1$ to facilitate hyperparameter tuning and prevent overfitting. \\
Besides, following the grid-invariance property discussed in Section \ref{sec: feature engineering}, this work utilizes a downsampled sequence of $N=$ 205 points per period, reduced from the original 1024 points. This strategy significantly mitigates computational overhead while preserving the capacity of the model to resolve complex underlying physical dynamics. In addition, with the validation data, the early stopping \cite{early_stop} strategy was implemented. The training process monitors the validation loss with a patience of 100 epoch. If no significant improvement (defined by a threshold $10^{-6}$ in this work) is observed within this window, the training is terminated, and the parameters corresponding to the lowest error are restored. This prevents the model from overfitting, ensuring the restoration of parameters corresponding to the optimal generalization state. Furthermore, this can significantly reduces unnecessary computational overhead \cite{early_stop}. By testing the model with different parameters on data of material 79, an initial baseline was established with a Pure FNO model, where an optimal set of hyperparameter was identified to balance predictive fidelity with computational efficiency (see Table~\ref{tab:fno_parameters}). These optimized parameters and sampling configuration remained constant in multi-scale Res-FNO to ensure a controlled performance comparison.

\begin{figure*}[!t]
    \centering
    % --- 第一行 ---
    \begin{minipage}[b]{0.45\textwidth}
        \centering
        \includegraphics[width=\textwidth]{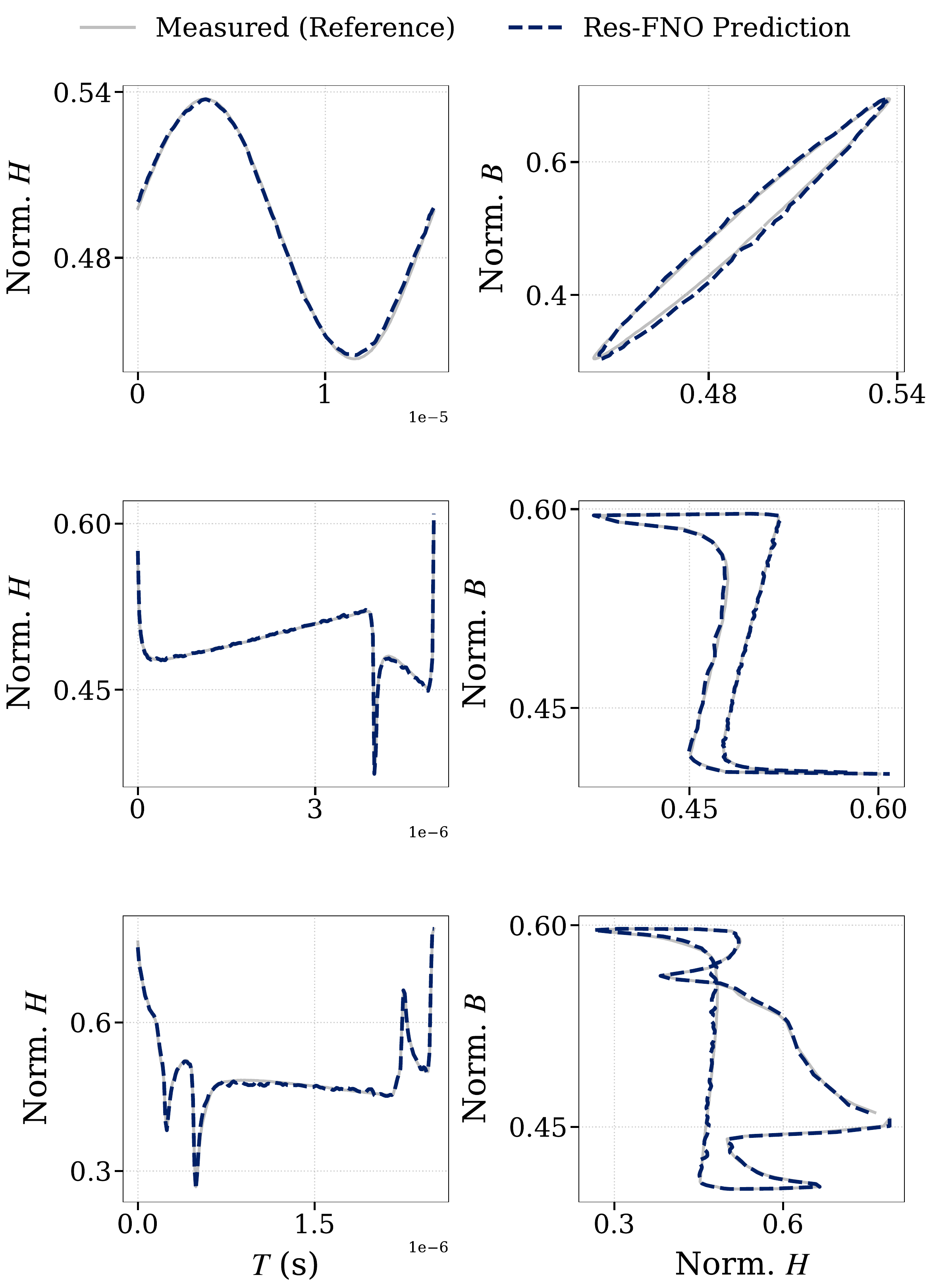}
        \subcaption{3C92}
        \label{fig:mat_A}
    \end{minipage}
    % \hfill % 撑开左右间距
    \begin{minipage}[b]{0.45\textwidth}
        \centering
        \includegraphics[width=\textwidth]{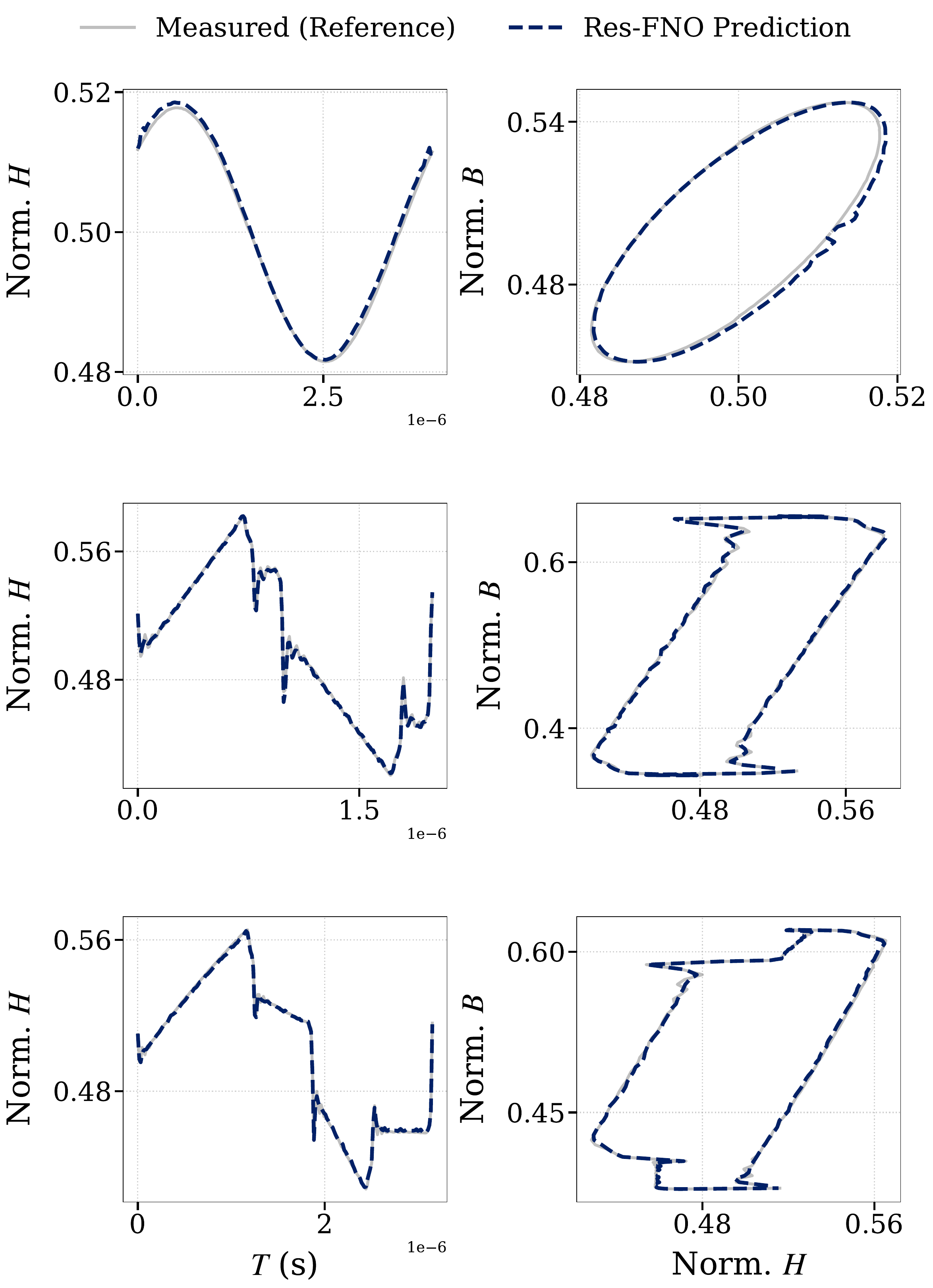} % 请确认此处文件名是否应改为另一个
        \subcaption{T37}
        \label{fig:mat_B}
    \end{minipage}

    \vspace{0.5cm} % 增加上下两行之间的间距

    % --- 第二行 ---
    \begin{minipage}[b]{0.45\textwidth}
        \centering
        \includegraphics[width=\textwidth]{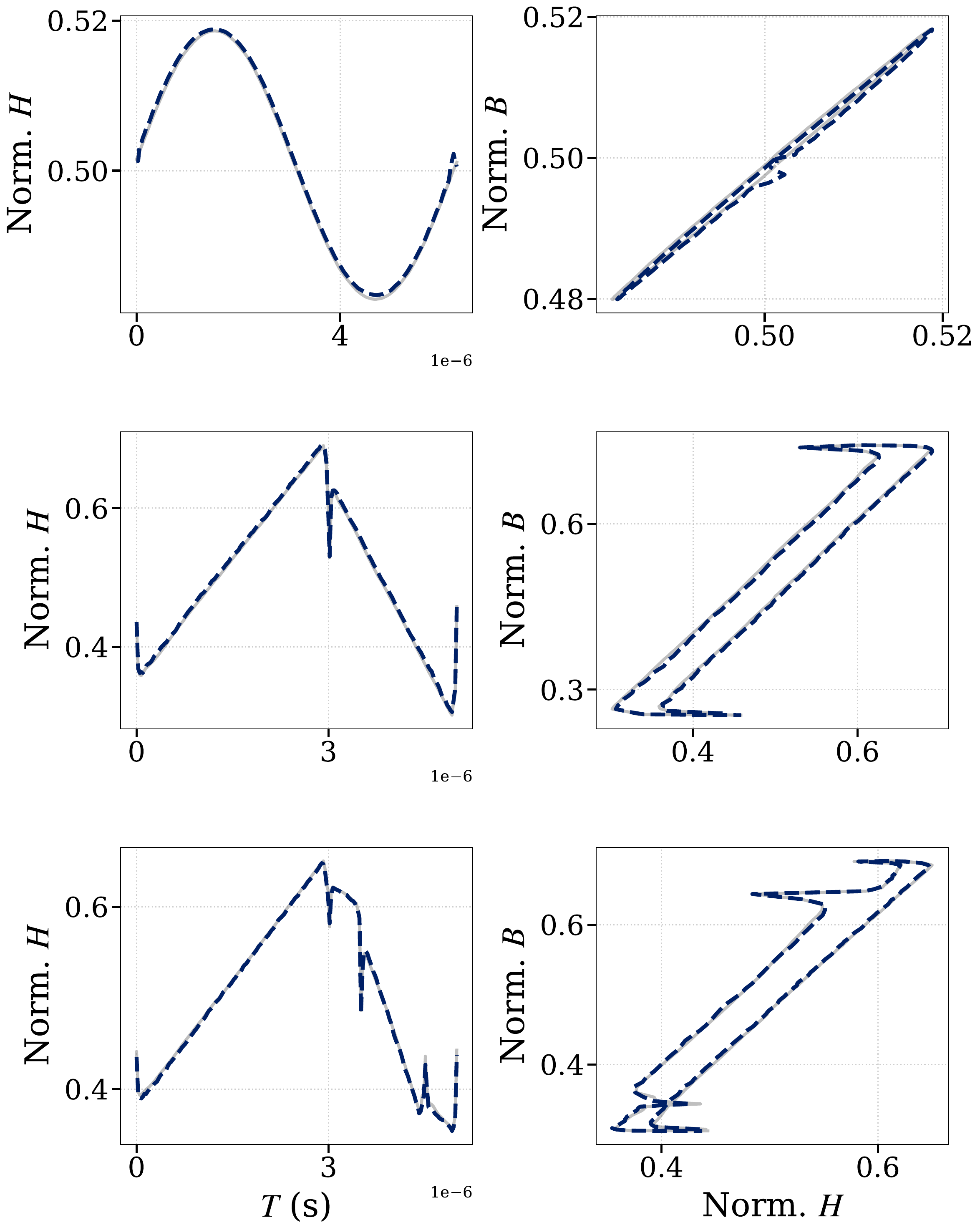}
        \subcaption{3C95}
        \label{fig:mat_C}
    \end{minipage}
    % \hfill
    \begin{minipage}[b]{0.45\textwidth}
        \centering
        \includegraphics[width=\textwidth]{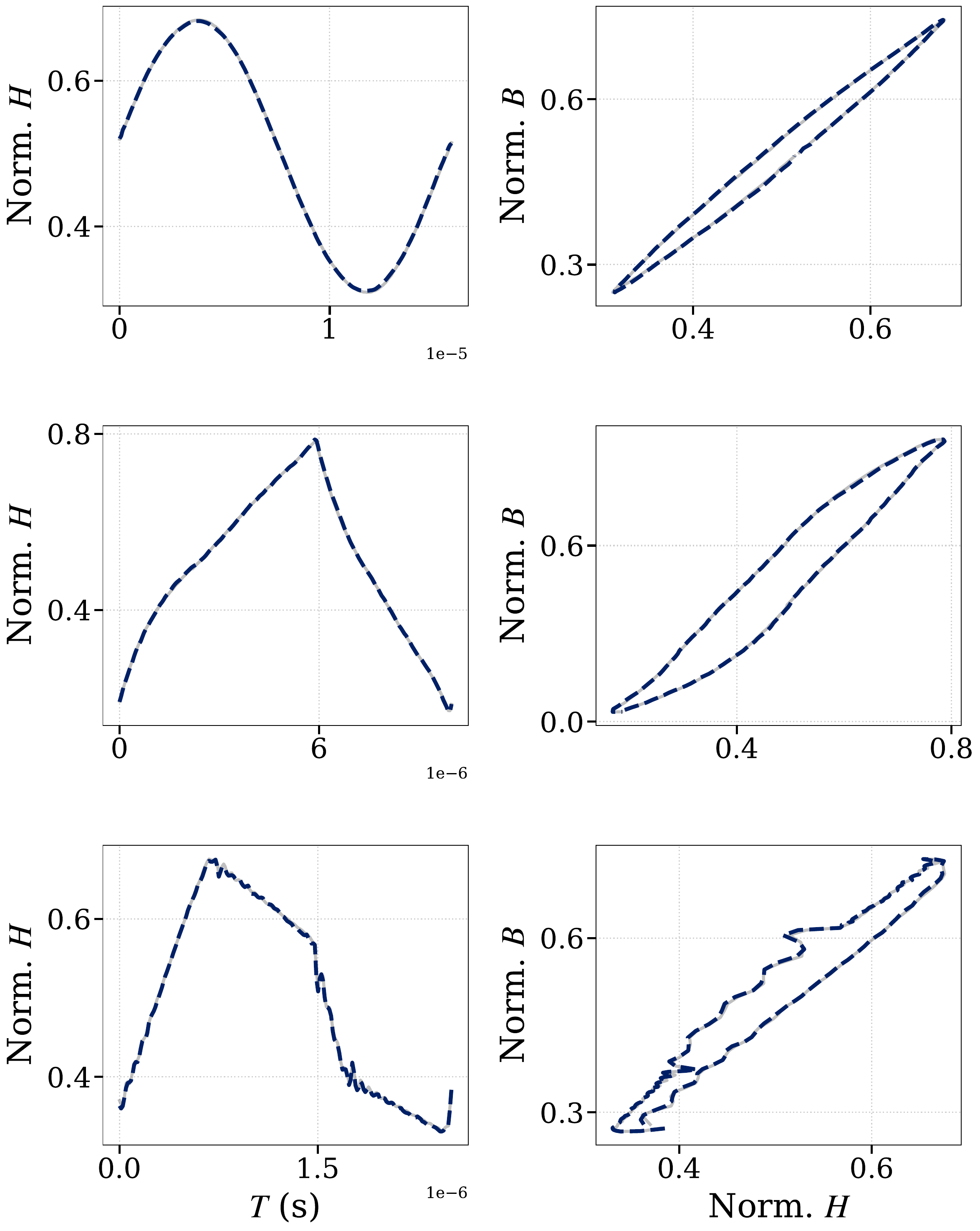}
        \subcaption{79}
        \label{fig:mat_D}
    \end{minipage}

    \caption{(a) Hysteresis loops of 3C92. (b) Hysteresis loops of T37. (c) Hysteresis loops of 3C95. (d) Hysteresis loops of ML95S.}
    \label{fig:Hysteresis_loops_ABCDE}
\end{figure*}

\begin{figure}
    \centering
    \includegraphics[width=0.9\linewidth]{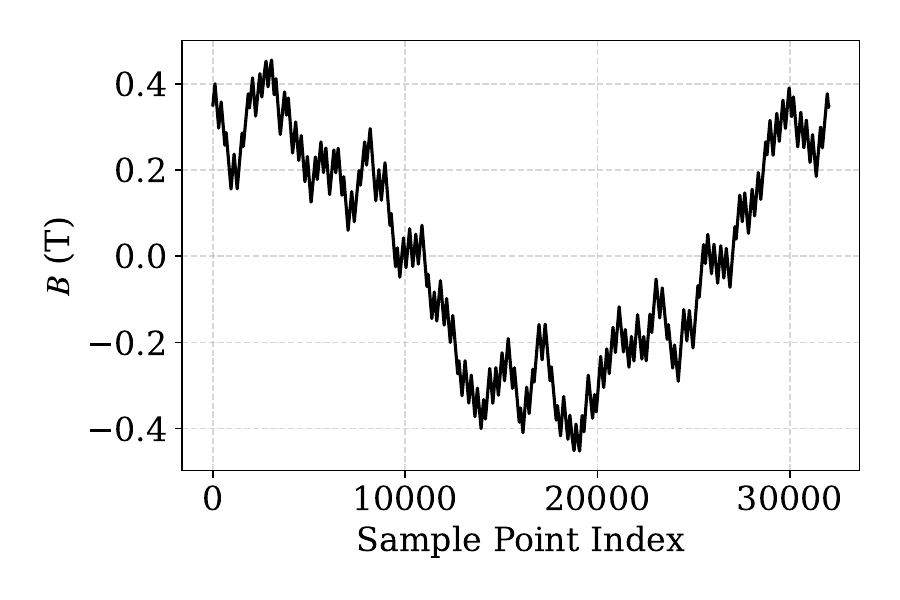}
    \caption{Exciting $B$ with oscillations.}
    \label{fig:exciting_B}
\end{figure}

\begin{table}[htbp]
\centering
\caption{Architectural parameters for the Pure FNO baseline model.}
\label{tab:fno_parameters}
\begin{tabular}{@{}lc@{}} % 修正为2列
\toprule
\textbf{Hyperparameter of Pure FNO} & \textbf{Value} \\ \midrule
Number of Fourier Layers ($L_{fno}$) & 2 \\
Number of Fourier Modes ($k$) & 48 \\
Hidden Dimension ($d_{model}$) & 64 \\
Activation Function & ReLU \\ 
Sequence Length ($N$) & 205 \\
Sequential Input Channels & 3 \\ 
% Training data size (Our model/Original data) & 521/579 ((Training : Test  as $9:1$) \\
% Test data size & 7298 \\ 
\bottomrule
\end{tabular}
\end{table}

\subsection{Ablation study on hybrid multi-scale Res-FNO components}
\label{sec:ablation study}
The proposed hybrid multi-scale architecture integrates two key design choices: enriched scalar inputs with the derivative $\frac{dB}{dt}$ and integrated ResNet structure. This ablation study systematically disentangles the contribution of each component.

\subsubsection{Influence of ResNet integrated FNO}
\label{sec:influence_of_Transformer}
After the preliminary study, 2 FNO blocks and 2 ResNet blocks, with the kernel size as 5 and 7 respectively are chosen, as illustrated in Table \ref{tab:resfno_hyperparameters}. To demonstrate the accuracy enhancement achieved by integrating ResNet into the FNO architecture, an ablation study was conducted on Material 79. As described in Section \ref{sec: problem statement}, the dataset was partitioned into training and validation sets with a ratio of $9:1$. So, there are respectively 521 and 7298 groups of data used for training and testing the model. The predictive performance of the various models is quantitatively assessed using two complementary metrics:

\begin{itemize}
    \item Sample-wise Normalized Root Mean Square Error (NRMSE): To evaluate the estimation accuracy across various magnetic excitation levels, the NRMSE is calculated for each individual magnetic field sequence as:
    \begin{equation}
        \label{eq:NRMSE}
        \text{NRMSE}_j = \frac{1}{H_{p,j}} \sqrt{\frac{1}{n} \sum_{i=1}^{n} (H_{j,i} - \hat{H}_{j,i})^2} \cdot 100%,
    \end{equation}
    where $j$ denotes the index of the test sample, $n$ is the number of time steps per period, and $H_{p,j} = \max|H_j(t)|$ represents the peak amplitude of the $j$-th measured sequence. This metric provides a scale-invariant measure, ensuring that the reconstruction fidelity is assessed consistently from linear regions to deep saturation.
    
    \item Coefficient of Determination ($R^2$): This metric assesses the goodness of fit for each predicted sequence $\hat{H}_j$ relative to its experimental ground truth $H_j$. The sample-wise $R^2$ score is defined as:
    \begin{equation}
        \label{eq:r_squared}
        R^2_j = 1 - \frac{\sum_{i=1}^{n} (H_{j,i} - \hat{H}_{j,i})^2}{\sum_{i=1}^{n} (H_{j,i} - \bar{H}_j)^2},
    \end{equation}
    where $\bar{H}_j$ denotes the arithmetic mean of the $j$-th experimental sequence. An $R^2$ value of $100\%$ indicates that the predicted curve perfectly coincides with the measured data. By calculating these metrics for every sample in the test set, a comprehensive statistical distribution of the performance of the model can be established.
\end{itemize}

\begin{table}[htbp]
\centering
\caption{Hyperparameter Configuration of Res-FNO for Different Datasets.}
\label{tab:resfno_hyperparameters}
\begin{tabular}{@{}lccc@{}}
\toprule
\textbf{Dataset / Materials} & \textbf{\makecell[c]{FNO \\ Blocks\\ ($n$)}} & \textbf{\makecell[c]{ResNet \\ Blocks\\ ($m$)}} & \textbf{\makecell[c]{Kernel \\ Sizes\\ ($k$)}} \\ 
\midrule
\makecell[l]{79, 3C92, T37, 3C95, ML95S} & 2 & 2 & \{5, 7\} \\ 
\midrule
3C90 (with Oscillations) & 2 & 3 & \{5, 7, 13\} \\ 
\bottomrule
\end{tabular}
\end{table}

As summarized in Table \ref{tab:ablation_study}, the proposed Res-FNO consistently outperforms the Pure FNO across all quantitative metrics. Specifically, the Res-FNO achieves a mean NRMSE of $1.87\%$, representing a significant error reduction compared to the $2.19\%$ produced by the Pure FNO baseline, with $14.8\%$ improvement in reconstruction fidelity. Furthermore, the average $R^2$ score is elevated from $99.79\%$ to $99.86\%$, indicating a superior goodness-of-fit to the experimental ground truth. The NRMSE distribution, illustrated in Fig. \ref{fig:NRMSE_comparison}, further confirms the enhanced robustness of the Res-FNO. Its error profile is more concentrated in the lower region with the leftward shift, suggesting higher reliability across diverse excitation conditions.\\ 
In addition, the integration of the residual structure is specifically designed to capture fine-grained transient details, such as the ringing effect. To provide a more intuitive comparison, some predicted magnetic field $H(t)$ and corresponding $B-H$ loops are plotted in Fig. \ref{fig:Hysteresis_loops_M79}. It is evident that the Res-FNO tracks the high-frequency oscillations caused by the ringing effect with much higher precision than the Pure FNO across all excitation types, demonstrating its superior ability to characterize complex magnetic dynamics.

\subsubsection{Efficacy of model inputs processing}
\label{sec:efficacy of multi-input processing}

Magnetic hysteresis characteristics vary significantly with temperature, frequency, and magnetic flux density. While incorporating these features as basic training inputs is essential, this study further introduces the time derivative $\frac{dB}{dt}$ into the sequential inputs, as discussed in Section \ref{sec: multi input processing}. the primary motivation is to counteract the ringing effect occurring during rapid transitions of $B$, as illustrated in Fig. \ref{fig:Ringing effect showing}. To evaluate the impact of this derivative information in input, a comparative model trained without $\frac{dB}{dt}$ was analyzed.

\begin{table}[htbp]
\centering
\caption{Performance comparison of different model architectures on the test dataset of Material 79.}
\label{tab:ablation_study}
\begin{tabular}{lcc}
\toprule
\textbf{Model Architecture} & \textbf{Average $R^2$ (\%)} & \textbf{Average NMSE (\%)} \\
\midrule
Pure FNO & 99.79 & 2.19 \\
Res-FNO (without $dB/dt$) & 99.83 & 2.00 \\
Res-FNO (Proposed) & 99.86 & 1.87 \\
\bottomrule
\end{tabular}
\end{table}

Using the same evaluation metrics, both the quantitative results in Table~\ref{tab:ablation_study} and the NRMSE distribution in Fig.~\ref{fig:NRMSE_comparison} demonstrate that the Res-FNO achieves better accuracy compared to the model without $\frac{dB}{dt}$. Qualitatively, the comparison in Fig.~\ref{fig:Hysteresis_loops_M79} further confirms that the proposed model effectively tracks the ringing effect better.

\subsection{Generalization and Statistical reliability analysis}
\label{sec:generalization study}
Having established the hybrid multi-scale Res-FNO as the optimal architecture through ablation studies on Material 79, we now evaluate its generalization capability across the broader materials: the others 4 materials in the second tier group of data provided in MagNet 2023 for the ringing effect modeling and, the data of 3C90 with minor loops provided in MagNet Challenge 2.
\subsubsection{Study on the materials with ringing effect}
\label{sec:first group of materials}

\begin{figure*}[ht]
    \centering
    \includegraphics[width=0.8\linewidth]{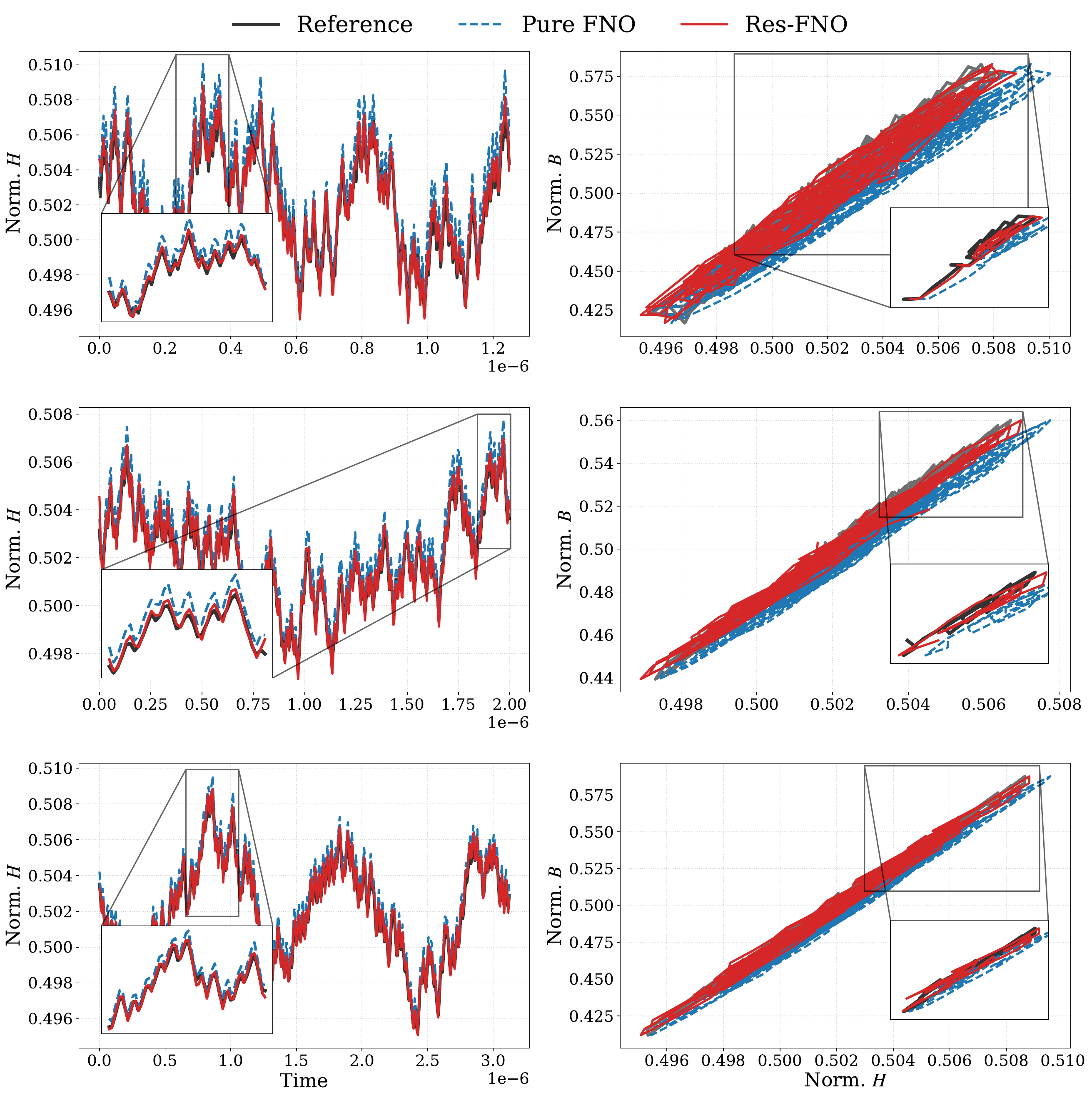}
    \caption{Comparison of predicted magnetic characteristics for using proposed Res-FNO and Pure FNO: (Left) Temporal waveforms of the magnetic field strength $H(t)$; (Right) Resulting $B$-$H$ hysteresis loops.}
    \label{fig:Hysteresis_loops_minor_loops}
\end{figure*}

\subsubsection{Study on the minor loops modeling}
\label{sec:second group of materials}

We evaluated the hybrid multi-scale Res-FNO on the other four materials: 3C92, T37, 3C95 and ML95S, which present distinct extrapolation challenges as outlined in \cite{Hardcore}. These materials test model robustness under data scarcity, domain shift, waveform sparsity, and extreme operating conditions. Considering the great generalization ability of proposed Res-FNO, we used only a sliced portion of the available training data for the training depending on the complexity of the materials. The data size are detailed in Table.~\ref{tab:data_info_4_materials}, where ``used/provided" means the data size provided originally in the challenge and the size we used for training our model. This sparse training scenario rigorously tests the ability of the model to extract generalizable physical principles from limited observations, mirroring practical situations where comprehensive characterization data for new materials may be scarce.

% \begin{table*}[htbp]
% \centering
% \caption{Dataset Statistics and Performance Summary for Hybrid PI-TFNO.}
% \label{tab:data_info_4_materials}
% \begin{tabular}{@{}lcccc@{}}
% \toprule
% \multirow{2.5}{*}{\textbf{Material}} & \multicolumn{2}{c}{\textbf{Dataset Size}} & \multicolumn{2}{c}{\textbf{Res-FNO Performance}} \\
% \cmidrule(lr){2-3} \cmidrule(lr){4-5}
%  & \textbf{Training (Used/Provided)} & \textbf{Testing} & \textbf{Average NRMSE (\%)} & \textbf{Average $R^2$ (\%)} \\ 
% \midrule
% \textbf{3C92}  & 2187/2432 & 7651 & 3.07 & 99.21 \\ 
% \textbf{T37} & 1332/7400 & 3172 & 2.43 & 99.68 \\ 
% \textbf{3C95}  & 964/5357  & 5357 & 2.10 & 99.78 \\ 
% \textbf{ML95S} & 905/2013  & 3738 & 0.96 & 99.97 \\ 
% \bottomrule
% \end{tabular}

% \end{table*}

\begin{table*}[htbp]
\centering
\caption{Dataset Statistics and Performance Summary of Res-FNO on the 3C92, T37, 3C95 and ML95S.}
\label{tab:data_info_4_materials}
\begin{tabular}{@{}lcccc@{}}
\toprule
\multirow{2.5}{*}{\textbf{Material}} & \multicolumn{2}{c}{\textbf{Dataset Size}} & \multicolumn{2}{c}{\textbf{Res-FNO Performance}} \\
\cmidrule(lr){2-3} \cmidrule(lr){4-5}
 & \textbf{Training (Used/Provided)} & \textbf{Testing} & \textbf{Average $R^2$ (\%)} & \textbf{Average NRMSE (\%)} \\ 
\midrule
\textbf{3C92}  & 2187/2432 & 7651 & 99.21 & 3.07 \\ 
\textbf{T37}   & 1332/7400 & 3172 & 99.68 & 2.43 \\ 
\textbf{3C95}  & 964/5357  & 5357 & 99.78 & 2.10 \\ 
\textbf{ML95S} & 905/2013  & 3738 & 99.97 & 0.96 \\ 
\bottomrule
\end{tabular}
\end{table*}

This cross-material validation is essential to demonstrate that the proposed architecture learns fundamental physical principles of hysteresis rather than overfitting to specific material characteristics. Table~\ref{tab:data_info_4_materials} and Figure~\ref{fig:Hysteresis_loops_ABCDE} summarize the performance of all these materials, which shows great accuracy and the generalization ability of the proposed Res-FNO.

In this section, the model is further implemented on Material 3C90, in which the exciting magnetic flux density $B(t)$ is not just sinusoidal or triangle, trapezoidal anymore but with the oscillation inside as shown in Fig. \ref{fig:exciting_B}, which is more realistic. The dataset spans seven discrete excitation frequencies ranging from 50 kHz to 800 kHz and three different temperature as 25, 50 and 70 $\degree C$. The minor loops characteristics differs a lot under different frequencies as shown in Fig. \ref{fig:minor_loops_showing}. Since the data was acquired over complete excitation cycles at a constant sampling rate, the raw sequence lengths varied inversely with the excitation frequency. To ensure a consistent input dimension for the neural network, we employed a linear resampling technique to align all temporal sequences to a fixed length of $2016$ points per cycle first. And then, all the data are sliced to only 504 points in each period. The reason why there is more time points in this one than 205 in the other materials is that there are more oscillations in this data as shown in Fig. \ref{fig:exciting_B}. In total, there are 17262 groups of data, to test the generalization ability of the model, only $10 \%$ of them (1726) groups of data randomly chosen from the data group are used for training the model, another $10 \%$ are used for validation. And the remaining $13810$ groups of data are used to test the accuracy of the model. As shown in Table \ref{tab:resfno_hyperparameters}, one more ResNet block with the kernel size as 13 is added since the complexity and the oscillation of the date.

% \begin{table}[htbp]
% \centering
% \caption{Performance Comparison between Pure-FNO and Res-FNO for 3C90 Ferrite Material.}
% \label{tab:3c90_comparison}
% \begin{tabular}{@{}lcc cc@{}}
% \toprule
% \multirow{2}{*}{\textbf{Model}} & \multicolumn{2}{c}{\textbf{Dataset Size}} & \multicolumn{2}{c}{\textbf{Performance Metrics}} \\
% \cmidrule(lr){2-3} \cmidrule(lr){4-5}
%  & \textbf{Training} & \textbf{Testing} & \textbf{Average $R^2$ (\%)} & \textbf{Average \\ NRMSE (\%)} \\ 
% \midrule
% Pure-FNO & \multirow{2}{*}{1726} & \multirow{2}{*}{13810} & 96.27 & 6.20 \\ 
% Res-FNO & & & 98.22 & 4.25 \\ 
% \bottomrule
% \end{tabular}

% \vspace{2pt}
% \begin{flushleft}
% \footnotesize  
% \textit{Note:} Both models were trained on the same 10\% subset (1726 samples) to evaluate their learning efficiency under limited data conditions.
% \end{flushleft}
% \end{table}

\begin{table}[htbp]
\centering
\caption{Performance Comparison between Pure-FNO and Res-FNO for 3C90 Ferrite Material.}
\label{tab:3c90_comparison}
\begin{tabular}{@{}lcccc@{}} % 修正了列定义中的多余空格
\toprule
\multirow{2}{*}{\textbf{Model}} & \multicolumn{2}{c}{\textbf{Dataset Size}} & \multicolumn{2}{c}{\textbf{Performance Metrics}} \\
\cmidrule(lr){2-3} \cmidrule(lr){4-5}
 & \textbf{Training} & \textbf{Testing} & \textbf{Average $R^2$ (\%)} & \textbf{\makecell[c]{Average \\ NRMSE (\%)}} \\ 
\midrule
Pure-FNO & \multirow{2}{*}{1726} & \multirow{2}{*}{13810} & 96.27 & 6.20 \\ 
Res-FNO  &                       &                        & 98.22 & 4.25 \\ 
\bottomrule
\end{tabular}
\vspace{2pt}
\begin{flushleft}
\footnotesize  
\textit{Note:} Both models were trained on the same 10\% subset (1726 samples) to evaluate their learning efficiency under limited data conditions.
\end{flushleft}
\end{table}

The performance of Res-FNO on 3C90 under oscillating excitations is analyzed and compared with the Pure FNO. As shown in Table \ref{tab:3c90_comparison}, with only 10\% (1726 samples) of the dataset used for training, Res-FNO achieves a superior $R^2$ of 98.22\% and reduces the NRMSE to 4.25\%, significantly outperforming the Pure-FNO. The qualitative comparison in Fig. \ref{fig:Hysteresis_loops_minor_loops} demonstrates the superior performance of the Res-FNO architecture in capturing complex magnetic dynamics. Unlike the Pure FNO model, which can not be accurate at high-frequency ripples in the $H(t)$ waveforms, the Res-FNO accurately tracks these sharp peaks. This temporal precision extends to the $B$-$H$ loops, where the Res-FNO effectively eliminates the phase lag and amplitude errors seen in the Pure FNO. As highlighted by the zoomed-in insets, the Res-FNO precisely follows the intricate trajectories of minor loops. Furthermore, achieving high accuracy on 13,810 unseen samples with limited training data underscores the robust generalization of the proposed multi-scale Res-FNO, proving that residual connections successfully bridge global spectral features with fine-grained local physical corrections.

% \afterpage{\clearpage} 
% 需要 \usepackage{afterpage}
% 或者直接用 \clearpage

\section{Conclusion}
\label{sec:conclusion}
To address the inherent spectral bias of the vanilla Fourier Neural Operator (FNO) in magnetic hysteresis modeling, this paper proposes a hybrid multi-scale ResNet-augmented FNO (Res-FNO) architecture for seq-to-seq hysteresis prediction. Specifically, a dedicated feature engineering approach is implemented to fuse scalar and sequential inputs using min-max normalization. A key contribution is the inclusion of the time derivative of magnetic flux density ($\frac{dB}{dt}$) as an auxiliary input, which is physically motivated by the fact that ringing effects and minor loop dynamics are highly sensitive to the rate of change of $B(t)$. Comprehensive ablation studies conducted on Material 79 validate the efficacy of this feature enhancement and demonstrate the superior capability of the integrated ResNet path in capturing high-frequency transients. Furthermore, the generalization performance of the proposed Res-FNO is rigorously evaluated using four distinct materials from the MagNet Challenge 2023, as well as Material 3C90 characterized by complex oscillations. Experimental results, quantified by NRMSE and $R^2$ metrics, along with the visualization of predicted $H$-field waveforms and hysteresis loops, consistently show that the ResNet-based refinement path significantly improves the modeling accuracy of minor loops. The high fidelity in reconstructing high-frequency oscillations and the robust performance across diverse materials prove that the proposed model is a powerful and reliable tool for the seq-to-seq modeling of complex hysteresis behaviors in power electronic applications.

\section*{Acknowledgment}
Ziqing Guo acknowledges support from China Scholarship Council (CSC), No. 202206280041, in the preparation of this manuscript.
\ifCLASSOPTIONcaptionsoff
  \newpage
\fi

\bibliographystyle{IEEEtran} % 设置参考文献样式为 IEEE 标准
\bibliography{reference}    % 这里的 references 对应你的 .bib 文件名（不含后缀）

\newpage

% \section{Biography Section}
% If you have an EPS/PDF photo (graphicx package needed), extra braces are
%  needed around the contents of the optional argument to biography to prevent
%  the LaTeX parser from getting confused when it sees the complicated
%  $\backslash${\tt{includegraphics}} command within an optional argument. (You can create
%  your own custom macro containing the $\backslash${\tt{includegraphics}} command to make things
%  simpler here.)
 
% \vspace{11pt}

% \bf{If you include a photo:}\vspace{-33pt}
% \begin{IEEEbiography}[{\includegraphics[width=1in,height=1.25in,clip,keepaspectratio]{fig1}}]{Michael Shell}
% Use $\backslash${\tt{begin\{IEEEbiography\}}} and then for the 1st argument use $\backslash${\tt{includegraphics}} to declare and link the author photo.
% Use the author name as the 3rd argument followed by the biography text.
% \end{IEEEbiography}

% \vspace{11pt}

% \bf{If you will not include a photo:}\vspace{-33pt}
% \begin{IEEEbiographynophoto}{John Doe}
% Use $\backslash${\tt{begin\{IEEEbiographynophoto\}}} and the author name as the argument followed by the biography text.
% \end{IEEEbiographynophoto}

\vfill
\end{document}